\RequirePackage{snapshot}
\documentclass [10pt, fancyhdr, twoside] {article}
\usepackage{float, graphicx, caption, amssymb}
\usepackage[usenames,dvipsnames]{color}
\usepackage{tabulary}
\usepackage [left=2.5cm, top=2.5cm, bottom=2.5cm, right=3cm] {geometry}  
\usepackage{fancyhdr}
\usepackage{xcolor}
\usepackage[scaled]{helvet}

\usepackage[left]{lineno}
\usepackage[yyyymmdd,hhmmss]{datetime}
\usepackage{authblk}

\usepackage{blindtext}

\newcounter {note}
\stepcounter{note}

\begin{document}
\title{\textbf{Novel Materials and Concepts for Next-Generation High Power Target Applications} \\ \vspace{0.2 in} \Large{A whitepaper for Snowmass 2022 from the AF-7 - Accelerator Technology R\&D Subgroup Targets and Sources} \vspace{0.3in}}

\author[1]{K. Ammigan}
\author[1]{S. Bidhar}
\author[1]{F. Pellemoine}
\author[1]{V.Pronskikh}
\author[1]{D. Pushka}
\author[1]{K. Yonehara}
\author[1]{R. Zwaska}
\author[2]{A. Couet}
\author[2]{M. Moorehead}
\author[3]{T. Ishida}
\author[3]{S. Makimura}
\author[4]{C. Densham}
\author[4]{M. Fitton}
\author[4]{E. Harvey-Fishenden}
\author[4]{T. Davenne}
\author[4]{D. Jenkins}
\author[4]{P. Loveridge}
\author[4]{J. O'Dell}
\author[4]{C. Rogers}
\author[4]{D. Wilcox}
\author[5]{M. Calviani}
\author[5]{S. Gilardoni}
\author[5]{F.-X. Nuiry}
\author[5]{A. Perillo-Marcone}
\author[6]{N. Bultman}
\author[6]{J. Gao}
\author[6]{M. Larmann}
\author[6]{M. LaVere}
\author[6]{W. Mittig}
\author[6]{M. Reaume}
\author[6]{J. Wei}
\author[6]{Y. Xu}
\author[7]{G. Muhrer}
\author[7]{T. Shea}
\author[7]{C. Thomas}
\author[7]{M. Wohlmuther}
\author[8]{E. Wakai}
\author[9]{C. Barbier}
\author[9]{B. Riemer}

\affil[1]{\normalsize{Fermi National Accelerator Laboratory, Batavia, IL 60510, USA}}
\affil[2]{\normalsize{University of Wisconsin-Madison, Madison, WI 53715, USA}}
\affil[3]{\normalsize{High Energy Accelerator Research Organization, Tsukuba, Japan}}
\affil[4]{\normalsize{Rutherford Appleton Laboratory, Didcot, Oxfordshire, UK}}
\affil[5]{\normalsize{European Laboratory for Particle Physics, Geneva, Switzerland}}
\affil[6]{\normalsize{Facility for Rare Isotope Beams, Michigan State University, MI 48824, USA}}
\affil[7]{\normalsize{European Spallation Source, Lund, Sweden}}
\affil[8]{\normalsize{Japan Atomic Energy Agency, Tokai, Japan}}
\affil[9]{\normalsize{Oak Ridge National Laboratory, Oak Ridge, TN 37831}}

\vspace{0.3in}
\date{\today}

\maketitle

%


\section{Introduction}

Novel beam-intercepting materials and targetry concepts are essential to improve the performance, reliability and operation lifetimes of next generation multi-megawatt (multi-MW) accelerator target facilities. The beam-intercepting materials and components must sustain an order-of-magnitude increase in particle beam intensities and are beyond the current state-of-the-art. With conventional materials and targetry technologies already limiting the scope of experiments \cite{Hylen2017,Kramer2014,Hasegawa2017}, it is crucial to investigate novel materials and concepts that will satisfy the requirements and maximize the physics benefits of future energy and intensity frontier experiments. High-power target innovation and novel materials R\&D are necessary to enable and ensure reliable operation of future target facilities.

In addition to the U.S. planned program for High Energy Physics (HEP) target facilities and upgrades such as the 2.4 MW Long-Baseline Neutrino Facility (LBNF) and muon-to-electron-conversion II experiment (Mu2e-II), international HEP program plans also feature challenging accelerator target facilities and related beam-intercepting devices. These include the Beam Dump Facility (BDF), HiLumi-LHC and Future Circular Collider (FCC) collimators at CERN, the T2K, COMET, and hadron experimental facilities at J-PARC, and beam-intercepting devices for the proposed International Linear Collider (ILC). Non-HEP accelerator target facilities, involving neutron sources, waste transmutation and nuclear physics applications also face similar challenges and include the 5-MW European Spallation Source, the Spallation Neutron Source (SNS) proton power upgrade project at Oak Ridge National Laboratory, the Facility for Rare Isotope Beams (FRIB) at Michigan State University, and the Material and Life Science Experimental Facility (MLF) at J-PARC. Meeting the demands of these accelerator target facilities and their planned beam power upgrades are of great concern to these institutions. Significant R\&D of novel targetry materials and concepts beyond the current state-of-the-art are therefore essential to enable these next-generation accelerator facilities.

\section{Challenges of beam-intercepting devices}

Beam-intercepting devices such as beam windows, beam dumps, collimators and particle production targets are designed to absorb the energy and power of the particle beam in order to produce and deliver the particles of interest to particular experiments. These devices are engineered to withstand the challenging beam-induced thermomechanical loads and optimized for physics performance. The continuous bombardment of these components by high-energy high-intensity pulses beams poses serious challenges to the operation and maintenance of target facilities. Beam-induced thermal shock and radiation damage effects in materials were identified as the leading cross-cutting challenges facing high-power target facilities \cite{Hurh2012}.

Thermal shock phenomena arise in beam-intercepting materials as a result of localized energy deposition by very short pulsed-beams (1-10 $\mu$s). The rapidly heated core volume of the material in the beam spot region expands but is constrained by the surrounding cooler target material. This condition creates a sudden localized region of compressive stress than propagates as stress waves through the material at sonic velocities after each beam pulse. If the initial dynamic stress exceeds the yield strength of the material, it will permanently deform and eventually fail. In addition, the cyclic loading environment from the pulsed beam progressively damages the material’s microstructure such that it can ultimately fail at stress levels that are actually lower than its failure strength (fatigue failure).

The bulk material properties change as a result of radiation damage in highly irradiated materials. Radiation damage disrupts the lattice structure of the material through the displacements of atoms, transmutation, and gas production after sustained particle beam bombardment. Irradiation-induced defects such as dislocation loops, point-defect clusters, fine-scale precipitates and voids that accumulate at the microstructural level ultimately affect the bulk properties of the material. Typical bulk property effects include embrittlement, hardening, swelling, reduction of thermal conductivity, and an increase in diffusion-dependent phenomena such as segregation of impurities, and phase transformation \cite{Kiselev2016}, all critical properties for the reliable and safe operation of beam-intercepting devices. As beam power and intensity increase, there is a pressing need to explore novel radiation damage and thermal shock tolerant beam-intercepting materials.

The removal of heat deposited into the beam-intercepting material upon interaction with the beam is another key challenge facing multi-MW beam-intercepting devices. Heat deposition in the material will increase as beam power increases, and therefore more effective cooling systems will be needed to safely operate these devices and to avoid compromising the physics performance. Liquid or gas cooling mediums are typically used to remove heat via forced convection from the boundary material. However, cooling efficiency from forced convection is largely dependent on the available cooling surface area which is usually driven by the desired physics performance (target size). Therefore, the limitation of forced convection cooling may impose significant constraints on the physics performance of future multi-MW devices. As a result, there is a real need to explore alternative advanced cooling technologies for next-generation beam-intercepting devices.

In addition to developing novel materials and understanding their behavior under high-energy high-intensity beam irradiation conditions, the development of novel targetry concepts and technologies are required to advance the design, fabrication and reliable operation of future multi-MW target facilities. Forward-looking concepts and technologies will push the current state-of-the-art and fully harness the benefits of novel materials to maximize the physics benefits of future experiments. The novel targetry concepts and technologies include pebble-bed, flowing and granular targets, rotating targets, liquid targets, specifically shaped targets to diffuse or damp stress waves, variable density targets taking advantage of additive manufacturing techniques, advanced cladding technologies, high heat-flux cooling and novel material coatings. 

The following sections of this whitepaper discuss the current and novel targetry materials, concepts and technologies that will need to be explored and developed further over the next few years in order to address the  challenges facing high-power beam-intercepting devices. Significant coordinated R\&D in these areas will be necessary to enable the safe and reliable operation of several future multi-MW target accelerator facilities.

\section{Novel targetry materials}
This section describes candidate novel materials that need to be explored and developed for specific beam-intercepting devices, capable of sustaining the increased beam power and intensity of future accelerator facilities.

\subsection{High-Entropy Alloys}
As opposed to the majority of relevant alloys used today (steels, aluminum alloys, titanium alloys, etc.), high-entropy alloys (HEAs) represent a fundamental departure from conventional metallurgy methodologies. Whereas conventional alloys are typically comprised of one or two principal elements with their properties tweaked by adding small alloying additions, HEAs consist of several principal elements often present in near equimolar quantities. The result is a material with a structure and properties that are not dictated primarily by a single element, but rather behave as an average of each primary constituent element. Beginning with two seminal papers in 2004, researchers Brian Cantor and Jien-Wei Yeh independently stumbled onto a trend in research which would grow exponentially over the next 15 years \cite{Cantor2004, Yeh2004}. Motivating their alloy design was a pursuit of maximizing the configurational entropy in expanded alloy systems in the hope that this increase in entropy would be sufficient to suppress the formation of intermetallic compounds, which are often deleterious to the mechanical properties of materials. Yeh is credited with coining the term “high-entropy alloy”, which he originally defined as equimolar alloys of five or more principal elements.

Over the years, HEAs have been demonstrated to exhibit a broad range of promising properties for both structural materials and functional materials including markedly higher yield strengths than Ni-based superalloys at high temperatures, fracture toughness comparable to the best cryogenic steels at low temperatures, high-temperature oxidation resistance, and superplastic behavior, as well as efficient catalysis of H$_2$ and CO$_2$ and the largest magnetocaloric effect experimentally observed in a material. But by far, the most relevant and intriguing property of HEAs for accelerator target applications is their microstructural response to irradiation damage. Many experimental studies have shown that certain HEA compositions outperform their less compositionally complex counterparts under irradiation \cite{Ullah2016, Yang2019, Tong2018, Atwani2019, Lu2016, Jin2016}, especially when comparing void swelling behavior. To understand the origins of the apparent irradiation tolerance exhibited by HEAs, many mechanisms have been proposed in the literature, the most commonly accepted being (i) a higher recombination during damage cascade, (ii) sluggish diffusion of point defects, (iii) broadening/overlapping defect migration energies, and (iv) sluggish diffusion of interstitial loops/clusters. While each of the above proposed mechanisms has been supported by some form of modeling and simulation, it still remains unclear which mechanism plays a dominant role in determining the performance of HEAs under irradiation. 

HEAs therefore offer a unique opportunity to explore a broader range of novel radiation-damage resistant alloy systems with functional properties specific to accelerator beam-intercepting applications. 

\subsection{Electrospun nanofiber materials}

Nanofiber materials offer promising applications as future multi-MW targets as they are intrinsically tolerant to both thermal shock and radiation damage. Recently Fermilab has engaged in the design and development of target materials with a sinuous microstructure consisting of ceramic nanofibers using a unique electrospinning set-up. Since the continuum is physically discretized at the microscale, issues such as thermal shock, thermal stress cycles and local heat accumulation can be mitigated \cite{Bidhar2021}. The microstructure consists of large number of randomly oriented one-dimensional nanofibers of less than hundred nanometers in diameter. Since the diameter of individual nanofibers are many orders of magnitude smaller than the beam spot size, there will not be any temperature gradient across the nanofiber cross-section. Moreover, there are a lot of gaps between the individual nanofibers which would prevent any kind of compressive stress waves to propagate. The large surface area of individual nanofibers and the porosity in the microstructure of the bulk nanofiber mat would also offer better cooling from the beam center. In-beam test of select nanofiber material specimens at CERN's HiRadMat facility revealed promising evidence of enhanced thermal shock resistance.

Owing to the nanopolycrystalline grains in individual nanofibers, it is hypothesized that they would also offer better resistance to radiation damage. The large number of grain boundaries and free surfaces would act as sinks to irradiation induced defects. The submicron diameter and ubiquitous grain boundaries of nanofibers reduce the mean free path for low-solubility transmutation products. Helium gas that forms due to high-energy proton beam interaction with the material can escape out of the material and hence avoid swelling which can cause the bulk material to crack. In order to evaluate their resistance against displacement damage, some of these nanofibers were irradiated under 1 MeV Kr++ beam at Argonne National lab IVEM Tandem facility. The samples received a 5 DPA-equivalent DPA in stainless steel. Selected area diffraction pattern (SADP) in TEM before and after irradiation shows no new peaks indicating phase stability of these nanofibers after irradiation. Comparison of d-spacing plots before and after shows there was also no change in d-spacing and peak location, implying no change in lattice parameters or any kind of amorphization. The dark and bright field TEM images also do not show any dislocation loops, clusters. However, more systematic studies are needed on the nanofiber material in order to correlate radiation damage and fluence effects caused by high-energy proton beam and low-energy heavy ion beam.

\subsection{SiC-coated graphite and SiC-SiC composites}

Graphite shows extremely high performance when used in proton beam target applications due to its thermal and mechanical properties, and chemical stability. However, graphite is also easily oxidized at high temperatures. If air is unexpectedly introduced into the primary beam line during a high-power beam operation, the graphite target can rapidly oxidized. And these graphite oxidation contaminants complicate the recovery procedures and downtimes. So, as an alternative to graphite, it is important to develop a material that is more resistant to oxidation.

Recently, work to investigate Silicon Carbide (SiC) coated graphite, an excellent candidate because of its good heat and high oxidation resistance, has begun. Under the Radiation Damage In Accelerator Target Environments (RaDIATE) collaboration \cite{RaDIATE}, a high-intensity proton beam exposure with 181 MeV energy was conducted at Brookhaven Linac Isotope Producer facility on various material specimens for accelerator target and beam window applications. The experiment included the SiC-coated graphite as a future target material in US and Japan high-intensity proton accelerator facilities. The radiation damage level of the SiC and graphite reached 0.24 and 0.05 DPA, respectively. Post Irradiation Examination testing of the irradiated specimens have been conducted at Pacific Northwest National Laboratory.

Nano-powder Infiltration and Transient Eutectoid (NITE) SiC/SiC composite, developed for the fusion and fission reactor at Muroran Institute of Technology, is another excellent candidate for target material because it is significantly denser than graphite \cite{Kohyama2011}. Higher efficiency of secondary-particles transport is estimated in some experiments by a Monte-Carlo simulation because the spatial volume of the source is reduced \cite{Makimura2020_1}. The NITE SiC/SiC exhibits much higher oxidation resistance than graphite \cite{Park2018} and exhibits a pseudo-ductile behavior, which enables it to withstand considerably higher stresses up to the fracture strength. The NITE SiC/SiC was irradiated and examined at CERN's HiRadMat facility as well \cite{Maestre2022}.

\subsection{Toughened Fine-Grained Recrystallized (TFGR) tungsten}

Tungsten (W) is a principal candidate as target material because of its high density and extremely high melting point. The use of W can provide 10 times higher brightness of muon/neutron than that of the current target materials \cite{Makimura2020_2}. While expected as the target material for the proton accelerator, W inherently has a critical disadvantage due ot its brittleness at around room temperature. The low temperature brittleness can be avoided by heavy plastic working, although the working can be sufficiently applied only to filaments or hot-rolled thin plates and its effect also depends on the working direction. However, even if the brittleness is alleviated, W exhibits significant embrittlement due to recrystallization that occurs when W is heated at and above the recrystallization temperature, which is almost one-third of the melting point. Moreover, it exhibits significant embrittlement by proton irradiation as well \cite{Linsmeier2017}. TFGR (Toughened, Fine Grained, Recrystallized) tungsten alloy, which was originally developed at Tohoku University and whose technology has been transferred to KEK and Metal Technology Co., LTD, has grain boundary reinforced nanostructures to overcome the embrittlement \cite{Kurishita2013,Makimura2021}. 

\subsection{Dual-phase titanium alloys}

Titanium alloys are one of the most suitable materials for accelerator beam windows due to their low density and high strength \cite{Ishida2018}. In J-PARC neutrino facility, thin domes of titanium 64 alloy (Ti-6Al-4V, Ti-64) cooled by helium are used as the primary beam window between the target station (helium) and the accelerator (vacuum)\cite{Ishida2019}, and a thin tube as an airtight container covering the entire graphite target \cite{Densham2009, Fishenden2019}. They have maintained stable operation so far without any serious failures. The same design concept will be applied to the LBNF target \cite{Papa2018, Wilcox2019}. In the FRIB rare ion beam facility, a rotating, water-cooled Ti-64 thin drum is used as the beam dump and a water-cooled Ti-64 sheet is also being considered for the beam window of the ILC main beam dump.

Ti-64 is a particular titanium alloy that has achieved a superior balance between strength and ductility by creating a fine equiaxed structure with a mixture of $\alpha$ (HCP) and $\beta$ (BCC) two phases through elemental addition and thermomechanical treatment. It can be used in high-temperature environments up to about 300 $^{\circ}$C and has excellent corrosion resistance. However, under neutron or proton beam irradiation, the alloy hardens significantly and loses almost all of its ductility after only 0.1 DPA \cite{Mansur2008}. It has been suggested that this may be due to the embrittlement of the irradiation-induced $\omega$-phase in the $\beta$-phase matrix, in addition to the hardening caused by the dense dislocation loops in the $\alpha$-phase matrix \cite{Ishida2020}. Therefore, under pulsed beam injection with higher flux and higher repetition rate which is expected in the future, the possibility of fatigue failure due to thermal shock cannot be ruled out and should be addressed.

The requirements for high-strength titanium alloys as next-generation beam window materials are to maintain enough strength and ductility even when irradiated at a few DPA under an operating temperature of about 200-300 $^{\circ}$C and maintain a service life of a few to several operational years. Based on the results of past research, the R\&D issues described below need to be addressed.

\begin{itemize}
\item So far, the beam window has been procured from bulk materials of Ti-64 available on the market, subjected to simple heat treatment such as stress relief, and machined without further microstructural controls. In terms of availability and machinability, Ti-64 has so far been the most suitable titanium alloy for beam windows. Meanwhile, two-phase titanium alloys are generally characterized by the possibility of rich microstructural control by thermomechanical treatment to achieve the desired mechanical properties, such as equiaxed, bi-modal, and needle-like microstructures. The improvement of irradiation resistance by microstructure control is worth investigating. The microstructure control in combination with the near-net-shape manufacturing technology by 3D printing, which has been developed remarkably in recent years, is also worth considering.

\item A wide variety of titanium alloys have been developed mainly in response to the requirements of the aerospace industry, and it is worth to compare their radiation damage resistance with that of Ti-64. A single metastable $\beta$ phase alloy Ti-15-333 has the potential to be a material with high radiation damage tolerance that does not undergo irradiation hardening at room temperature \cite{Ishida2018_1}.  This owes to the nano-sized dense precursor of the $\omega$ phase in the $\beta$ phase matrix, acting as very effective sink sites to absorb irradiation defects. Heat treatment to maintain the irradiation resistance up to high temperatures is being investigated. A similar effect would be expected for other metastable $\beta$ alloys that have been precipitation-strengthened, e.g. TIMET's beta21S. Precipitation strengthening has also been applied to some high-temperature near-$\alpha$ titanium alloys used in aircraft engines, which may have high resistance to radiation damage for the same reason as metastable $\beta$ alloys, e.g. DAIDO’s DAT-54.

\item In the J-PARC beam window, chemical corrosion marks, which may be caused by impurities in the helium atmosphere and beam current, have been observed on the target station side. As one of the methods to improve the corrosion resistance of titanium alloy, the coating treatment of TiN and TiAlN by Physical Vapor Deposition (not CVD) may be effective, and the evaluation of its irradiation resistance and thermal shock resistance is necessary.

\item The effect of embrittlement and swelling due to hydrogen and helium, which are spallation products of proton beam irradiation, on mechanical properties should be evaluated. This is necessary not only for titanium alloys but also for all target and beam window materials.
\end{itemize}

\subsection{Advanced graphitic materials}

Novel advanced graphitic materials are also essential for targetry and in general for beam intercepting devices, owing to their high temperature resistance, relatively low coefficient of thermal expansion and low density. Different grades and properties of graphitic based absorber exists. Isostatically-pressed graphite is a widely employed and cost-effective solution, which have been validated with beam impact exceeding 10 kJ/cm$^3$. This material also presents excellent performance under radiation damage and present, due to its crystal lattice configuration, good annealing capabilities. 2D carbon/carbon is another class of high-performance graphitic-based material, due to the conformation of the material structure. These materials have been regularly employed since the start of the Large Hadron Collider for highly demanding applications and its use has been further expanded in the framework of the LHC Injectors Upgrade (LIU) project, where a new class of 3D carbon/carbon materials are employed. 2D CC composites have large dimensions in the plane (directions 1 and 2) compared with their thickness (direction 3), and their mechanical properties are generally weaker in direction 3. In comparison, 3D CC composites can have relatively large dimensions in the third or Z direction, together with interesting mechanical properties \cite{Nuiry2019}. In specific cases, electrical conductivity plays an important part, to reduce the overall machine impedance and increase beam stability. For this purpose, a family of novel graphite-based composites reinforced with a dispersion of molybdenum carbide particles \cite{Valenzuela2018}, with very high thermal and electrical properties, have been developed at CERN in collaboration with the European industry, and further R\&D is necessary.

All the graphitic materials mentioned above have been tested at the HiRadMat facility at CERN, to demonstrate their capabilities to withstanding the extremely challenging operational conditions encountered in the accelerator chain.

\section{Novel targetry concepts and technologies}
This section covers some of the key novel targetry concepts and technnologies that need to be explored and optimized to enable and support future multi-MW accelerator target facilities.

\subsection{Rotating, flowing and circulating targets}

The Facility for Rare Isotope Beams (FRIB) at Michigan State University will use projectile fragmentation and induced in-flight fission of heavy-ion primary beams at energies of 200 MeV/u and up to 400 kW beam power for experiments in nuclear physics, nuclear astrophysics, and fundamental science \cite{Wei2012}. One of the major challenges of the FRIB project is the design and integration of the high power target systems. To achieve required high resolution of the fragment separator, the beam spot on the production target needs to be on the order of 1 mm size with about 100 kW beam power deposited in the target, leading to a power density of 20-60 MW/cm$^3$ and heavy ion induced radiation damage of the material. A rotating solid carbon disk was selected as the technical baseline concept for all primary beams up to Uranium. 

The FRIB target uses a multi-slice design with a diameter of about 30 cm and rotational speed of about 5000 RPM to provide efficient heat dissipation by thermal radiation and to minimize temperature variations as well as the associated thermal mechanical stress and fatigue. This allows the beam spot temperature in the solid carbon material to maintain a maximum of 1900 $^{\circ}$C. To match the expected duration of typical experiments at FRIB, a lifetime of the order of two weeks would be required. And recently, the FRIB team successfully commissioned a krypton-86 primary beam using the FRIB rotating target \cite{Wei_soon}. 

As FRIB beam power ramps up to 400 kW, the target disk module will be replaced by single-slice and multi-slice targets. The motor drives the target wheel up to 5000 RPM and the power deposited into the solid target can be radioactively cooled by surrounding water cooled heat exchanger and cover plates. A better understanding of the radiation damage of graphite under heavy ions is critical for FRIB beam power ramping up. Since the energy loss of the heavy ions in the material is much higher than that from protons, the radiation damage of the FRIB target, which could lead to a short lifetime, is of great concern. It has been observed that permanent damage occurred in the carbon stripper foils with heavy ion beams \cite{Marti2010}. With thicker graphite material and annealing temperature, it is expected that the graphite targets can satisfy FRIB target requirements but R\&D activities of graphite material study under heavy ion beam irradiation are needed for future developments.

With rotational speed up to 5000 RPM and high beam intensity, the bearing used in the system needs to be vacuum compatible, high radiation resistant and the lubricant needs to have high temperature tolerance. An induction heating system is being developed to investigate bearings made of various materials and temperature tolerance and radiation resistance of different vacuum compatible lubricants. Besides, collaboration with industrial providers on the design of 3D-printed targetry bearings made of the new filament materials is ongoing. \\

The European Spallation Source is designed to deliver a high neutron flux, produced by 5 MW, 3 ms proton pulse repeating at 14 Hz, which will enable unprecedented neutron base science \cite{Garoby2017}. One of the many challenges brought by the high power density is the design of the spallation target. The spallation material not only has to withstand the extremely high power density, but has to last long enough to support reliable operation with close to 100$\%$ availability. The materials not directly exposed to the proton beam are exposed to extremely high dose of hadron, neutrons and gamma rays particles. 

The conceptual design for the operation of the target aims at controlling and reducing the power density of the proton beam, together with reducing the heat and radiation damage to the spallation material. The first aspect is achieved by sweeping the beam position across the target area during the passage of the pulse. The second aspect aims at lowering average dose on the spallation material. This is achieved with the rotating target concept.

It is clear that the condition for the spallation target (tungsten) of 5 MW the material are submitted to very high stress. Not to mentioned that an uncontrolled beam will invariably lead to damage of the materials, the high radiation dose associated with high temperature will lead to change of material properties and potential high damage and rupture of materials. The process of spallation will produce 56 neutrons for each of the 10$^{15}$ proton per pulse. The average current density on target is 53 $\mu$A/cm$^2$, so the proton fluence on spallation material per year is 6 $\times$ 10$^{21}$ protons per cm$^2$. For each pulse, the amount of energy for a 2-GeV beam is 357 kJ. Any material under these conditions will experience extremely high stresses. Thermal effect will heat materials in ms to 100s of degrees, generating stresses in the 50 to 100 MPa range. Materials on the target wheel will be exposed to the beam once every 2.57 s, which leaves time for cooling, but also leads to cyclic stresses.

In the tungsten bricks, the stress is close to 50 MPa, and it will go through close to 40 $\times$ 10$^6$ stress cycles. The total dose in the tungsten for the 5 years lifetime of the target is over 10 DPA. Studies on Tungsten showed that within a month of 5 MW operation, the tungsten will be totally brittle \cite{Habainy2019}. However, the stress level remains far from the rupture point. Therefore, the material is expected to survive 5 years in the ESS beam, but some uncertainty remains due to lack of data for this material in these condition of temperature and irradiation dose. The target wheel is one of the main components of the spallation source. However, there are other components highly critical to ensure a high availability for neutron science. The other components are the proton beam window, moderator-reflector system, the target wheel driving system, the containment and shielding components, the beam diagnostics components. All these components are exposed to the high flux of particles coming from the spallation target. In addition, the proton beam window, which confines the target environment from the accelerator environment, and the beam diagnostics components intercept the beam and are submitted to the same constraints and stresses as the target wheel and spallation material. R\&D activities for the design and materials behavior under irradiation of these components are key in enabling the full potential of ESS. The rotating target concept proposed for position-production for the proposed ILC also faces very similar challenges. \\

There are three possible target designs that are being studied at Fermilab for the Mu2e-II pion-production target. A rotation target design consists of a set of tungsten rods revolving around an axis alternately by switching the rods exposed to the beam. The advantage of this design would be to distribute the heat deposited and radiation damage uniformly over several rods. A constraint however is to fit the target system into the Mu2e baseline Heat and Radiation Sheild (HRS) 25-cm radius inner bore. A fixed granular target design consists of a matrix of granular tungsten to be cooled by the flow of gaseous helium. Such a design would fit the existing HRS dimensions but its high radiation damage would likely require frequent replacement of the target. A third design under consideration is the conveyor target. Spherical tungstens or carbon elements will be supplied to a pipe, moved to the beam interaction area, and then removed from HRS for cooling and replacement (when necessary). This design would occupy a relatively small space (consistent with the HRS). Helium gas could be used for both cooling and moving elements inside the conveyor's pipe. Radiation damage can also be distributed among a large number of replaceable elements. The design is however technically complex and will require prototyping and extensive testing.

Preliminary MARS15 Monte Carlo and ANSYS Finite Element Analysis (FEA) simulations indicated that the conveyor design will require about 285 spherical target elements to be situated in the pipe inside the HRS inner bore while 28 elements in the case of carbon or 11 elements in the case of tungsten, will be located in the beam interaction region at any particular time. The elements will be moving inside the pipe with a velocity of ~10 elements/s (0.1 cm/s). Further analyses and R\&D are ongoing and will be required to realize this proposed target design and optimize the target performance.\\

Flowing liquid metals have also been successfully used as high-power neutron-production targets \cite{Bauer01,Mason06,Arai09}. The technology allows higher time-averaged beam power on target without degrading neutron brightness incurred with increasing coolant volume in solid, stationary targets. However, pulsed beam liquid-metal targets can suffer from pressure waves that drive cavitation in the liquid and fatigue of the vessel that contains the liquid metal. Cavitation-induced erosion damage has been a severe issue for the target vessel container \cite{Haines05, Riemer2008}. The cavitation erosion and fatigue can gradually degrade the structural integrity of the target vessel and lead to premature failure that affects the operational reliability of the facility. Significant and sustained R\&D has been carried out to mitigate the effects of pressure waves by injecting helium gas bubbles into the liquid metal. The approach has been highly successful in allowing the targets to operate at high beam powers for more extended periods \cite{Jiang22, Clintock, Kogawa2017, Naoe2021, Naoe2020}. The success demonstrates the payoff in dedicated target R\&D investment. Nevertheless, as beam power continues to increase the issues of cavitation erosion and fatigue may appear again. It is instrumental in continuing the R\&D work in this area. \\

A muon collider has a combination of requirements that are well beyond the limit of any existing target technology. A high-Z target is required to be suspended within the bore of a high field solenoid and subject to the high pulsed power density of a multi-MW proton beam. Flowing granular tungsten pneumatically conveyed within a pipe is being proposed and explored as an alternative to the current baseline technology proposal of an open mercury jet. The High Power Targets group at RAL has developed a fluidised tungsten powder target technology which combines some of the advantages of a liquid metal with those of a solid. The granular material flowing within a pipe is expected to be able to withstand extremely intense pulsed beam powers without the cracking or radiation damage limitations of solid targets, and without the cavitation issues associated with liquid targets \cite{Riemer2008}. Moreover, its disruption speed inside a gaseous helium atmosphere has been measured to be of the order of a few m/s \cite{Chari}. However, a fluidised powder target introduces new challenges, such as achieving reliable circulation and continuous stable horizontal dense phase flow, managing heat dissipation, mitigating radiation damage and erosion of the containing pipework and beam windows, as well as ensuring reliable diagnostics and controls for the powder handling processes.

An offline test rig was built at the Rutherford Appleton Laboratory (RAL) in order to demonstrate the feasibility of pneumatic fluidization and conveyance of powdered tungsten \cite{Caretta2008,Densham2009_1}. The rig can fluidize and lift sub-250 micron powder using suction and eject it in solid dense phase as a coherent open jet or contained pipe flow with a bulk fraction of c.50$\%$. Air was used for the test rig but helium is proposed for actual target applications due to its favorable heat transfer properties and to minimize radiological issues. Contained flow is proposed as being most suitable for use as a particle-production target \cite{Davies2010}. Subsequent developments have enabled the rig to continuously recirculate the powder, providing an uninterrupted stream of target material. 

The response of a static open trough of tungsten powder to a high energy proton beam was investigated in 2012 and 2015 at the HiRadMat facility at CERN \cite{Eft2011}. Eruption velocities from the free surface, due to the ionisation of the grains by the proton beam, were much lower than for liquid mercury subjected to the same energy density \cite{Caretta2014,Caretta2018,Davenne2018}, although for the latter this effect was significantly reduced by the capture solenoid magnetic field. It is anticipated that the beam-induced disruption should be considerably lower for a powder contained inside a tube although this would need to be demonstrated in a future HiRadMat experiment. Overall, a contained powder target system is expected to be considerably less damaging to its surroundings than an open liquid mercury jet, and also less problematic from a radiological point of view.

The fluidized tungsten powder concept shows promise as a target for a Muon collider or future high-intensity CLFV experiments \cite{Aoki2020}, but further development will be required to demonstrate its suitability for use in an operating facility. The fluidized tungsten powder concept currently operates entirely by timed “batch” processes involving a number of pneumatically operated sliding gate valves. An operating facility would ideally eliminate such moving parts. Careful selection of the pipework and beam window material will be required (e.g. SiC-SiC composite). Bespoke designs for high-erosion regions such as bends may be required, and long-term erosion measurements essential to demonstrate that the required target lifetimes of months or years can be achieved. Measurements of the heat transfer between the flowing tungsten powder and the surrounding containment tube would also be desirable. A future on-line experiment at HiRadMat is intended to investigate the effect of an intense pulsed proton beam on tungsten powder contained within a tube, using laser doppler velocimetry to measure any stresses transmitted from the granular material to the pipe wall. In addition to this practical work, an engineering feasibility study will investigate how to integrate the complete tungsten powder system within a capture solenoid for a muon collider target station. This will require input from a comprehensive physics design study, which will use simulation codes to calculate the predicted particle production rates and deposited energy densities in order to select the best geometry layout and beam parameters that will optimize the performance of the muon collider within the expected engineering constraints.

\subsection{Advanced cladding materials and technologies}
The European Laboratory for Particle Physics (CERN) have expanded the use of diffusion bonding assisted by Hot Isostatic Pressing (HIP) at CERN for beam intercepting devices systems, and applied for the LHC Injector Upgrade (LIU) Project \cite{Damerau2014} as well as for the Physics Beyond Colliders initiative \cite{Jaeckel2018}, in a similar fashion as the development done at existing spallation sources where Ta-cladded pure W are regularly employed for neutron production. Bonding dissimilar material with the HIP methods enhances the heat transfer coefficient between the different materials and therefore increases the capabilities to cope with power dissipation. 

Cuprous materials (such as Cu-OFE, CuCr1Zr and ODS alloys such as Glidcop or Discup) have been bonded via HIP with stainless steel. The technology was already developed for fusion reactors \cite{Marois1996} but CERN have further expanded the technique for very large components (up to 2.5 meters), required for the fabrication of the LIU-SPS internal beam dump (so called TIDVG5 \cite{Pianese2018,Pianese2021}), which is operational since 2021. The technology has a large potential for dumps and absorbers for a variety of different facilities, not only for proton driven, but also for electron or photon driven (synchrotron light) facilities, where the dissipated power requirements could be in excess of 100 kW. 

Refractory metals are also widely employed in laboratories worldwide for secondary beam production, such as for neutron production \cite{Thomason2018} or for proposed beam dump experiments \cite{Ahdida2019}, owing to their reduced nuclear interaction length and relatively low neutron inelastic cross-section. Nevertheless, directly cooling with water is not possible due to the relatively high hydrogen embrittlement. Usually tungsten is cladded via HIP with pure Ta, with good results \cite{Nelson2012}. For the proposed CERN’s Beam Dump Facility Project \cite{Ahdida2020}, CERN has developed advanced techniques to clad pure W as well as Mo-alloys (such as TZM) with pure Ta as well as, for the first time, Ta2.5W \cite{Lopez2019,Busom2019}. Beam irradiation of prototype target have been successfully executed during 2018 \cite{Lopez2019_1} with post irradiation examination (PIE) to be further expanded during 2021 and 2022. For high power facilities, decay heat on Ta and Ta-alloys may pose safety concerns: for this reason, within the framework of BDF, other cladding materials are being studied, including zircalloy or other Nb-alloys such as C103, which still possess excellent corrosion resistance capabilities, formability and resistance to high temperatures. C103 is a complex refractory metal, consisting of mainly Nb with addition of 10 wt$\%$ of hafnium and 1 wt$\%$ of titanium. These R\&D techniques will be beneficial for future Hidden Sector experiments \cite{Ahdida2019} as well as for other neutron production facilities, such as the ORNL SNS Second Target Station project \cite{ORNL}. The R\&D should be complemented with instrumented in-beam tests \cite{Lopez2019_1}, both at fast and slow extraction facilities, as well as with post irradiation examination (PIE) techniques, in order to validate the simulation packages and ensure the bonding quality (mechanical and thermal) after irradiation. \\

For neutron production at the ISIS Neutron and Muon Facility run by UKRI STFC at the Rutherford Appleton Laboratory, solid-plate water-cooled targets have been used for 38 years and monolithic water-cooled targets for 14 years.  Initially the plates in the solid-plate targets were made of depleted uranium clad in zircaloy, but these uranium targets suffered from premature failure due to radiation-induced swelling. Consequently, in 1995, the plates were switched to tantalum entirely. And then in 2001, the solid-plate targets were designed with tungsten plates clad in 1-2 mm of tantalum. Since 2001 the TS1 (Target Station 1) tantalum-clad tungsten solid-plate targets, have run to the present day successfully with no apparent cladding issues (four targets in total).  All of these targets were taken out of service simply because a few of the many thermocouples measuring plate temperatures eventually failed. TS1 is currently being upgraded and will include the next evolution of this tantalum-clad tungsten solid-plate target. In contrast, the TS2 (Target Station 2) tantalum-clad monolithic tungsten targets, which have run from 2008 to the present, have had a much shorter operational life than originally planned due to activation of the target water-cooling circuit.  The activation by spallation products of tungsten and tantalum becomes very evident after approximately 18 months of operation of the target, and clearly indicates a breach in the tantalum cladding which is exposing the tungsten to the cooling water.  Neutronically these targets are still performing as required, but they have to be replaced every two years or so to ensure that we keep radiation dose rates in our water plant to a sensible level for maintenance operations.

The current focus is to understand the mechanism for the cladding breach in the TS2 targets, and of course to find a solution to this problem.  There are a number of projects looking at stresses and fatigue processes in the tantalum during operation and at residual stresses produced during the process of HIP-ing the tantalum cladding on to the tungsten core. Offline research is ongoing to investigate the potential for erosion and corrosion of the tantalum cladding, as well as to carry out to carry out manufacturing process and QA improvement programs on materials supply and particularly on the EB welds which are used to join the tantalum cladding components together prior to the HIP-ing process.

In addition, new collaborative studies into radiation damage effects in the tungsten cores have prompted a re-visit of the engineering design parameters for the target materials. In particular, apparent reductions in thermal conductivity and in tensile strength may need to be considered in greater detail in the designs, especially for higher-power higher-intensity targets planned for future ISIS upgrades. There is also a continuing concern about the tantalum contribution to decay heat, especially in an unexpected loss-of-coolant scenario.  Thinning of the HIP-ed tantalum cladding is not a preferred option, particularly for the TS2 targets where there is not only potential concerns about water erosion and corrosion but also potential concerns about grain growth after EB welding which could lead to worryingly low numbers of grains in a thinner cladding layer and perhaps lead to accelerated cracking in the cladding.

For the future, efforts are underway to continue investigations into a better understanding of our current target materials (tungsten and tantalum), but also to explore the possibility of using TZM as a replacement cladding material. The immediate benefit of using the latter material would be a lower decay heat contribution which would help the loss-of-coolant scenario mitigation planning considerably. The development of advanced cladding materials technologies is fundamental in designing reliable and robust beam intercepting devices for planned upgrades and future Intensity and Energy Frontier facilities. 

\subsection{Target geometry and composition optimization}
Designing a target system is a tedious process because the optimization is done with various iterations to maximize the physics output for the experiment and to minimize the operational risk like the target fail. Usually, a Monte Carlo (MC) numerical simulation and a Finite Element Analysis (FEA) are used to evaluate the physics performance and to examine the mechanical stress on the target system in several iterations. The number of iterations exponentially grows by increasing the number of variances. To simplify the optimization, the designed target shape is usually monolithic, and the target material is single substance. However, a fine dimensional tuning of the target shape and finding the best mixing of substances for the target material are required to increase the performance efficiency of multi-MW target systems. 

The use of Machine Learning based on the Bayesian optimization is proposed for optimizing the target system. The Bayesian optimization will evaluate the previous result via the Gaussian process recurrence and set a new variance. This method is widely used in the material science field to find new materials/alloys. Combining additional simulator to design the novel target material at the atomic level may be used as well. This simulation effort will build a database server and utilize the available High Power Computing (HPC) for the target system optimization. 

\subsection{High-heat flux cooling}
Alternative advanced cooling technologies will need to be explored in order to address the challenges of heat removal from future multi-MW beam-intercepting devices. Unconventional heat transfer techniques, such as controlled boiling or even flowing (liquid or granular) targets, where heat-flux cooling capacity can be significantly increased, is an area that needs to be explored and developed. 

High heat-flux cooling techniques have been investigated in the past and continue to be researched within the fusion/fission communities, as well as in some accelerator target facilities. Methods to utilize boiling in a stable fashion (hyper-vapotron) have been developed for plasma-facing components in ITER, and can be applied to accelerator target applications. Radiative cooling is another cooling technique that will be used to cool the Mu2e tungsten target. Exploring other alternatives to forced convection cooling techniques is an area of research that should be addressed and critical in enabling future multi-MW target facilities. 

\section{Conclusion}
R\&D of novel targetry materials, concepts and technologies is necessary to enable and optimize the operation of ambitious future accelerator target facilities. The high power targetry community, primarily through the RaDIATE collaboration \cite{RaDIATE}, has been working on several R\&D projects to address the material  challenges of high power targets since 2012. This globally coordinated R\&D effort is essential to advance the current state-of-the-art in targetry. Increased funding and support for this research over the next several years is pivotal in achieving the objectives of the future research programs and experiments.

\end{document}